\pdfoutput=1
\documentclass[11 pt]{article}  

\usepackage[margin=1in]{geometry}
\usepackage{graphicx}
\usepackage{url}
\usepackage{times}
\usepackage{epstopdf}
\usepackage{algorithmic}
\usepackage{amsmath} 
\usepackage{amssymb}  
\usepackage{balance}

\newcommand{\E}{{\bf E}}

\newcommand{\eat}[1]{}

\makeatletter
\def\@begintheorem#1#2{\sl \trivlist \item[\hskip \labelsep{\bf #1\ #2:}]}
\def\@opargbegintheorem#1#2#3{\sl \trivlist
      \item[\hskip \labelsep{\bf #1\ #2\ #3:}]}
\makeatother

\title{\LARGE \bf
Queues with Small Advice
}

\author{Michael Mitzenmacher\thanks{School~of Engineering and Applied Sciences, Harvard University.  
Supported in part by NSF grants CCF-1563710 and CCF-1535795.  This is an arxiv draft, to be submitted and subject to changes.}}

\begin{document}

\maketitle
\thispagestyle{empty}
\pagestyle{empty}

\begin{abstract}
Motivated by recent work on scheduling with predicted job sizes, we
consider the performance of scheduling algorithms with minimal advice,
namely a single bit.  Besides demonstrating the power of very limited
advice, such schemes are quite natural.  In the prediction setting,
one bit of advice can be used to model a simple prediction as to
whether a job is ``large'' or ``small''; that is, whether a job is
above or below a given threshold.  Further, one-bit advice schemes can
correspond to mechanisms that tell whether to put a job at the front
or the back for the queue, a limitation which may be useful in many
implementation settings.  Finally, queues with a single bit of advice
have a simple enough state that they can be analyzed in the limiting
mean-field analysis framework for the power of two choices.  Our work
follows in the path of recent work by showing that even small amounts
of even possibly inaccurate information can greatly improve scheduling
performance.
\end{abstract}

\section{Introduction}
\label{sec:intro}
\subsection{Motivation}
\label{subsec:motivation}
In queueing settings where the required service time for a job is
known, strategies that take advantage of that information, such as
Shortest Job First (SJF) or Shortest Remaining Processing Time (SRPT)
can yield significant performance improvements over blind strategies
such as First In First Out (FIFO).  However, exact knowledge of the
service times is a great deal to ask for in practice.  Here we
consider the setting where one is given much more limited information.
Specifically, we consider the case where, for each job, a queue gets
only one bit of information, or advice, regarding the job size.  

While a one bit limitation may seem unusual, there are both
theoretical and practical motivations for such a study.  Online
algorithms with small amounts of optimal advice has been a subject of
study in the theoretical literature (see, e.g., the
survey \cite{boyar2016online}); such work highlights the potential for
additional information to improve performance.  Considering the case
of just one bit of information is an interesting limiting case.
Further, one-bit advice can naturally correspond to informing whether
a job should be placed at the front or the back of the queue; for some
queue implementations, such as in hardware or other highly constrained
settings, one may desire this simplicity over more complicated data
structures for managing job placement in the queue.

However, as a more concrete practical motivation, recently researchers have
studied queues with {\em predicted} service times, rather than exact
service times, where such predictions might naturally be provided by a
machine learning
algorithm \cite{DCM,mitz2019scheduling,mitzenmacher2019supermarket,oursurvey,
purohit2018improving,WiermanNuyens}.
Indeed, the queueing setting is one natural example of an expanding
line of work where predictions can be used to improve algorithms,
particularly in scheduling (e.g., \cite{hsu2018learning,TCFLIS,DBLP:conf/icml/LykourisV18,oursurvey,purohit2018improving}).
Our setting here of one-bit predictions can model a natural setting
where the prediction corresponds to whether a job's service time is
believed to be above or below a fixed threshold.  Such predictions may
be simpler to implement or more accurate than schemes that attempt to
provide a prediction of the exact service time.

For single-queue settings, our work uses standard queueing theoretic
analysis techniques.  Here we generally follow the (folklore) approach
of using Kleinrock's Conservation Law to derive formulae for the
conditional waiting time of a job according to its service time; this
approach dates back to at least the work of O'Donovan \cite{donovan},
from whose framework and notation we borrow.  The derivations can also
be readily obtained using the analysis of priority systems, following
the framework presented in for example \cite{harchol2013performance}.
The goal here is not to suggest new methods of analysis, but instead:
\begin{itemize}
\item show how the problem of scheduling with limited predicted information
can naturally be analyzed;
\item demonstrate how even limited advice and predictions can provide large performance gains;  and
\item show some interesting derivations for the special case we refer to as exponential predictions.
\end{itemize}

We also examine one-bit predictions schemes with large numbers of
queues using the power of two choices.  Here each arrival chooses the
better of two randomly selected queues (or more generally from $d$
randomly selected queues) from a large system of (homogenous) queues.
This study shares many of the same motivations as for single queues;
moreover, it may offer a first step to some open questions in the
area, such as analyzing the power of two choices when using Shortest
Remaining Processing Time or related schemes (see
e.g. \cite{mitzenmacher2019supermarket}).

Finally, more generally, we believe this work also highlights some
aspects of using machine learning predictions that may provide
guidance for the design of machine learning prediction settings.  For
example, we see that some predictions may be much more important than
others; in queueing settings, it seems generally much more important
to identify long jobs correctly than short jobs, as long jobs will
block many other jobs from service.  

\section{Single Queues and One-Bit Threshold Schemes}
\label{sec:thresh}

\subsection{Notation and Model}
\label{subsec:notation}

We consider M/G/1 queueing systems, with arrival rate $\lambda$ and
where the processing times are independently sampled according to the
cumulative distribution $F(x)$ with corresponding density $f(x)$.  
We follow some of the notation from \cite{donovan}.
We assume the expected service time has
been scaled so the mean service time is 1 (that is, $\E[F]=1$). 
Note $\E[F^2]$ is the second moment for the
service time.  We further let 
\begin{align*}
V & = \frac{\lambda \E[F^2]}{2}
\end{align*}
be the expected remaining service time of the job being served 
at the time of a random arrival.
We also let 
\begin{align*}
\rho(t) & = \lambda \int_{0}^t xf(x) dx
\end{align*}
be the rate at which load is added to the queue
from jobs with service time at most $t$, and 
correspondingly 
\begin{align*}
\rho(t) & = \lambda \int_{\infty}^t xf(x) dx = \lambda.
\end{align*}
\eat{
We also let 
$\lambda(t) = \lambda F(t)$, 
\begin{align*}
m(t) & = \int_{0}^t \frac{x f(x)}{F(t)} dx ,
\end{align*}
and $\rho(t) = \lambda(t) m(t) = \lambda \int_{0}^t xf(x) dx$.
For convenience we use $\rho$ to represents the rate at
which load is added to the system;  here $\rho = 
\lambda \int_{0}^\infty xf(x) dx = \lambda$.
}

\subsection{The Conservation Law}
\label{subsec:conservation}

As described in \cite{donovan}, Kleinrock's Conservation Law says that for a queue
with Poisson arrivals satisfying basic assumptions (such as the
queue is busy whenever there are jobs in the system), the expected load $L$ on the system at a random 
time point (e.g., in the stationary distribution), satisfies
\begin{align*}
L (1-\rho) & = V,
\end{align*}
where again $V$ is the expected load due to the job in service and
$\rho$ is the total rate at which load is added to the system.
The law allows simple derivations of
conditional expected waiting times, by looking at appropriate
subsystems of jobs. 

\subsection{Analysis of One-Bit Threshold Schemes}
\label{sec:analthresh}

We consider the case of an advisor that provides a single bit of
advice per job.  Specifically we consider the strategy where the
advice bit is 0 if the job's service time is less than some threshold
$T$, and 1 otherwise.  The job is placed at the front of the queue if
the advice bit is 0, and at the back of the queue otherwise.  We
consider preemptive and non-preemptive queues, where in the preemptive
case a job placed at the front will preempt the job currently
receiving service.  We later generalize the one bit of advice to
prediction-based systems, where the prediction is 
whether the service time for the job is larger or smaller than the
threshold.

\subsection{The non-preemptive system}
\label{subsec:nopreempt}

We first consider jobs arriving jobs with service time at most $T$.
Here we do not require the conservation rule; such a job is placed at the
front of the queue, although it has to wait for the job, if any, in
service to complete.  Further, any additional jobs of service time at
most $T$ that arrive before this job starts service is placed ahead of
the arriving job being considered.  We denote the expected waiting
time for a job with service time $t$, by which we mean the time spent by an incoming job in the
stationary distribution waiting before starting to obtain service, by
$W_1(t)$.  We denote the expected sojourn time, by which we mean the
entire time spent by an incoming job in the system, by $S_1(t)$.

The expected time an arriving job has to wait for an existing
job being processed is $V$.  It follows from standard busy
period analysis that additional incoming jobs increase
the expected waiting time by a factor of $\rho(T)$, and so
\begin{align*}
W_1(t) & = \frac{V}{1-\rho(T)}.
\end{align*}
The expected sojourn time for such jobs is thus
\begin{align*}
S_1(t) & = \frac{V}{1-\rho(T)} + t.
\end{align*}

For jobs with service time $t$ larger than $T$, we consider the
subsystem of all jobs, and use the notation $W_2(t)$ and $S_2(t)$
for the corresponding quantities.    
In this setting we have 
\begin{align*}
L & = \frac{V}{1-\rho} 
\end{align*}
from the conservation law.
For any job with service time larger than $T$, any new job 
with service time at most $T$ that arrives will be placed
ahead of of this job until it is served.  
Hence 
\begin{align*}
W_2(t) & =  \frac{L}{1-\rho(T)} \\
       & =  \frac{V}{(1-\rho)(1-\rho(T))},
\end{align*}
and
\begin{align*}
S_2(t) & =  \frac{V}{(1-\rho)(1-\rho(T))} + t.  
\end{align*}

For a given service distribution $F$ and threshold
$T$, we have the total expected waiting time $W$ in the system is
\begin{align*}
W & = \frac{V F(T)}{1-\rho(T)} +  \frac{V(1-F(T))}{(1-\rho)(1-\rho(T))} \\
& = \frac{V (1- F(T)\rho)}{(1-\rho)(1-\rho(T))}.
\end{align*}

The expected sojourn time $S$ satisfies $S = W +1$. 
Minimizing $W$ (or $S$) can be accomplished numerically. 

\eat{
** Was expecting a single threshold, but that does not appear to be
the case.
Note that the numerator is decreasing in $T$ while the denominator is
also decreasing in $T$.  

By taking the derivative of the waiting time with respect to $T$,
we find the waiting time is minimized when
\begin{align*}
(1-\rho(T)) (-f(T) \rho) + (1-F(T) \rho)\rho T f(T) & = & 0.  
\end{align*}

\rho(T) = 1 - (1-F(T) \rho)\rho T.  
1 - rho T - rho^2 TF(T)
- rho -rho^2 F(T) 
inc = 
} 

As an example we discuss through this work, for exponentially distributed service times, $V=\lambda$, $F(T) = 1 - e^{-T}$, $\rho = \lambda$, and
$\rho(T) = (\lambda)(1-(T+1)e^{-T})$.  
We find the expected sojourn time for this case, which we refer
to as $S_{e,n}$, is then 
\begin{align*}
S_{e,n} & = \frac{\lambda(1-\lambda+\lambda e^{-T})}{(1-\lambda)(1-(\lambda(1-(T+1)e^{-T})))} + 1 \\ 
& = \frac{1 - \lambda + \lambda (T+1)e^{-T} -  \lambda^2 T e^{-T}}{(1-\lambda)(1-(\lambda(1-(T+1)e^{-T})))} + 1.
\end{align*}
Taking the derivative, we find the optimal $T$ value occurs
when 
\begin{align*}
\frac{1}{\lambda} -1 & = \frac{e^{-T}}{T-1},  
\end{align*}
or equivalently we seek $T$ that satisfies
\begin{align*}
\lambda & = \frac{T-1}{e^{-T}+T-1},  
\end{align*}
In particular, as $\lambda$ goes to $1$, the optimal $T$ increases to
infinity, and as $\lambda$ goes to 0, the optimal $T$ goes to 1.
It is perhaps worth noting that a threshold $T$ of 4 corresponds to a 
$\lambda$ larger than $0.99$;  that is, in this case, we do not see very
large thresholds even under high load.  

As another example, we consider service distributions following the
Weibull distribution with cumulative distribution $F(x) = 1-e^{-\sqrt{2x}}$.
The Weibull distribution is heavy-tailed; while the average service
time of this distribution remains 1, the second moment is 6, so there
are many more very long jobs as compared to the exponential
distribution.  Weibull distributions are commonly used for queueing
simulations, as heavy-tailed service time distributions are more
realistic for many scenarios.

For this Weibull distribution, $V=3\lambda$ and $\rho(T)
= \lambda(1-e^{-\sqrt{2T}}(T+\sqrt{2T}+1))$ are
computed easily.
The expected sojourn time in this case, which we
denote by $S_{w,n}$, is then given by
\begin{align*}
S_{w,n} & = \frac{3\lambda(1-\lambda+\lambda e^{-\sqrt{2T}})}{(1-\lambda)(1-(\lambda(1-e^{-\sqrt{2T}}(T+\sqrt{2T}+1))))} + 1.
\end{align*}

\subsection{The preemptive system}
\label{subsec:preempt}

We first consider jobs arriving jobs with service time at most $T$.
Again, here we do not require the conservation rule; such a job is
placed at the front of the queue, and any additional job of
service time at most $T$ that arrive before this job starts service is
placed ahead of the arriving job being considered.  We use
$W_1(t)$ and $S_1(t)$ as before.

Clearly $W_1(t) = 0$. 
However, we also consider the effect of preemptions.
Since any job of size at most $T$ will preempt the job,
the expected sojourn time is
\begin{align*}
S_1(t) & = \frac{t}{1-\rho(T)}.
\end{align*}

For jobs with service times larger than $T$, we consider the
subsystem of all jobs, and again use the notation $W_2(t)$ and $S_2(t)$
for the corresponding quantities.    

In this setting we have 
\begin{align*}
L & = \frac{V}{1-\rho} 
\end{align*}
from the conservation law.
While waiting any job of service time at most $T$
is placed ahead of any job of size greater than $T$, 
so again
\begin{align*}
W_2(t) & = \frac{L}{1-\rho(T)} \\
       & = \frac{V}{(1-\rho)(1-\rho(T))}.
\end{align*}
Because of the preemption, the expected time from the
start of service until finishing service increases
to $t/(1-\rho(T))$, and so
\begin{align*}
S_2(t) & = \frac{V}{(1-\rho)(1-\rho(T))} + \frac{t}{1-\rho(T)}.
\end{align*}

In this case the total expected waiting time is
\begin{align*}
W & = \frac{V(1-F(T))}{(1-\rho)(1-\rho(T))},
\end{align*}
and the total expected sojourn time is 
\begin{align*}
S 
  & = \frac{V(1-F(T))+1-\rho}{(1-\rho)(1-\rho(T))}.
\end{align*}

For exponentially distributed service times, 
we find the expected sojourn time in this case, which we refer to
as $S_{e,p}$ is then 
\begin{align*}
S_{e,p} & = \frac{1 - \lambda + \lambda e^{-T} }{(1-\lambda)(1-(\lambda(1-(T+1)e^{-T})))}.
\end{align*}

One can readily that $S_{e,p} < S_{e,n}$ for any value of $T$;
indeed, 
\begin{align*}
\lambda S_{e,p} & = S_{e,n} - 1,
\end{align*}
so 
\begin{align*}
S_{e,n} - S_{e,p} & =  1 -(1-\lambda) S_{e,p} > 0.
\end{align*}
Also, the optimal value of
$T$ is again given by
\begin{align*}
\frac{1}{\lambda} -1 & = \frac{e^{-T}}{T-1}.  
\end{align*}

For the Weibull distribution, we have the corresponding expression
\begin{align*}
S_{w,p} & = \frac{1 - \lambda + 3\lambda e^{-\sqrt{2T}}}{(1-\lambda)(1-(\lambda(1-e^{-\sqrt{2T}}(T+\sqrt{2T}+1))))}.
\end{align*}

\section{Adding Predictions}
\label{sec:predict}

We consider a simple model where the probability of a misprediction
for a given item depends only on its service time, independent of
other jobs and other considerations.  Specifically, we suppose we have
a desired threshold $T$, and our prediction is simply our best guess
as to whether a job's service time is larger or less than $T$.  We
define $g_T(x)$ be the probability that a job of size $x$ is predicted
to be less than $T$.  While one can imagine more complex prediction
models, this model is quite natural, and is useful for examining the
potential power of predictions.

\subsection{The non-preemptive system}
\label{subsec:pnopreempt}

To deal with the predictions, we now let
\begin{align*}
Q(T) & = \int_{0}^\infty f(x)g_T(x) dx, 
\end{align*}
and 
\begin{align*}
\rho'(T) & = \lambda  \int_{0}^\infty xf(x)g_T(x) dx. 
\end{align*}
Here $\rho'(T)$ can be interpreted as the rate load arrives to 
the system from jobs with {\em predicted service time} at most $T$,
and similarly $Q(T)$ is the fraction of jobbs predicted to have 
service time at most $T$.

We first consider arriving jobs with predicted service time at most $T$
and actual service time $t$.
Following the same reasoning we have previously used, the waiting time ${W'}_t$ for such jobs is given by 
\begin{align*}
{W'}_1(t) & = \frac{V}{1-\rho'(T)}.
\end{align*}
For jobs with predicted service time greater than $T$,
\begin{align*}
{W'}_2(t) & = \frac{V}{(1-\rho)(1-\rho'(T))}.
\end{align*}
The total expected waiting time per job is therfore given by
\begin{align*}
W' & = \frac{V \int_{0}^\infty f(x)g_T(x) dx }{1-\rho'(T)}
+
\frac{V \int_{0}^\infty f(x)(1-g_T(x))dx }{(1-\rho)(1-\rho'(T))} \\
& = 
\frac{V ( 1- \rho \int_{0}^\infty f(x)g_T(x) dx ) }{(1-\rho)(1-\rho'(T))} \\
& = 
\frac{V ( 1- \rho Q(T)) }{(1-\rho)(1-\rho'(T))}.
\end{align*}

In particular, we see that the only changes from the setting without the prediction
is that in the $1- F(T)\rho$ term in the numerator, the $F(T)$ has been replaced
by the more complex integral expression $Q(T)$, and similarly the    
$1-\rho(T)$ term in the denominator has become $1-\rho'(T)$.

A model suggested in \cite{mitz2019scheduling,mitzenmacher2019supermarket} considers the setting where a prediction for a job with
service time $z$ is itself exponentially distributed with mean $z$;
we refer to this as the exponential prediction model.  While not necessarily
realistic, this model often allows for mathematical derivations, and provides
a useful starting point for considering the effects of predictions.
With this model,
\begin{align*}
g_T(x) & = 1 - e^{-(T/x)},   
\end{align*}
and hence 
\begin{align*}
Q(T) & = \int_{0}^\infty f(x)g_T(x) dx \\
& = 1 - \int_{0}^\infty e^{-x-(T/x)} dx \\
& = 1-2\sqrt{T} K_1(2\sqrt{T}),
\end{align*}
where $K_1$ is a modified Bessel function of the second kind.
Also 
\begin{align*}
\rho'(T) & =  \lambda  \int_{0}^\infty (xe^{-x} - xe^{-x-(T/x)}) dx \\ 
& =  \lambda (1- 2T K_2(2\sqrt{T})),   
\end{align*}
where $K_2$ is a modified Bessel function of the second kind
(with a different parameter).

The expected sojourn time for this case, which we refer to as 
$S_{e^*,n}$, is then
\begin{align*}
S_{e^*,n} & = \frac{\lambda (1-\lambda(1-2\sqrt{T} K_1(2\sqrt{T})))}
{(1-\lambda)(1-\lambda (1 - 2T K_2(2\sqrt{T})))} + 1.     
\end{align*}
There does not appear to be a simple form for the derivative of this
expression that allows us to write a simple form for the optimal value
fo $T$, although it can be found numerically.

We can perform similar calculations for our Weibull distribution.  
Here we have 
\begin{align*}
Q(T) & = 1 - \int_{0}^\infty \frac{1}{sqrt 2x} e^{-\sqrt{2x}-(T/x)} dx \\
& = 1 - \sqrt{\frac{T}{2\pi}} G_{0,3}^{3,0}\left(\frac{T}{2}~|~-\frac{1}{2},0,\frac{1}{2} \right ),
\end{align*}
where here $G$ is the Meijer $G$-function.  

Similarly 
\begin{align*}
\rho'(T) & = \lambda  \int_{0}^\infty \sqrt{\frac{x}{2}} \left (e^{-\sqrt{2x}} - e^{-\sqrt{2x}-(T/x)} \right ) dx \\ 
& = \lambda \left ( 1 - \sqrt{\frac{T^3}{2\pi}} G_{0,3}^{3,0}\left(\frac{T}{2}~|~-\frac{3}{2},0,\frac{1}{2} \right ) \right ),
\end{align*}
where again $G$ is the Meijer $G$-function.  

The expected sojourn time for this case, which we refer to as 
$S_{w^*,n}$, is then
\begin{align*}
S_{w^*,n} & = \frac{3\lambda \left (1 - \lambda \left (1 - \sqrt{\frac{T}{2\pi}} G_{0,3}^{3,0}\left(\frac{T}{2}~|~-\frac{1}{2},0,\frac{1}{2} \right ) \right ) \right )}
{(1-\lambda)  \left ( 1- \lambda \left ( 1 - \sqrt{\frac{T^3}{2\pi}} G_{0,3}^{3,0}\left(\frac{T}{2}~|~-\frac{3}{2},0,\frac{1}{2} \right ) \right ) \right )} + 1.
\end{align*}

\subsection{The preemptive system}
\label{subsec:ppreempt}

For the preemptive system, 
again let
\begin{align*}
Q(T) & = \int_{0}^\infty f(x)g_T(x) dx, 
\end{align*}
and 
\begin{align*}
\rho'(T) & = \lambda  \int_{0}^\infty xf(x)g_T(x) dx.
\end{align*}
We first consider jobs arriving jobs with {\em predicted} service time at most $T$
and actual service time $t$
Such jobs will have no waiting time, but their expected sojourn time is 
\begin{align*}
S_1(t) & = \frac{t}{1-\rho'(T)}.
\end{align*}

For jobs with predicted service time greater than or equal to $T$, 
following the same reasoning as in the case without predictions, we 
have
\begin{align*}
W_2(t) & = \frac{V}{(1-\rho)(1-\rho'(T))},
\end{align*}
and
\begin{align*}
S_2(t) & = \frac{V}{(1-\rho)(1-\rho'(T))} + \frac{t}{1-\rho'(T)}.
\end{align*}
We therefore find the total expected waiting time is
\begin{align*}
W & = \frac{V(1-Q(T))}{(1-\rho)(1-\rho'(T))},
\end{align*}
and the total expected sojourn time is 
\begin{align*}
S & = \frac{V(1-Q(T))}{(1-\rho)(1-\rho'(T))} + \frac{1}{1-\rho'(T)} \\
  & = \frac{V(1-Q(T))+1-\rho}{(1-\rho)(1-\rho'(T))}.
\end{align*}

For the exponential prediction model, 
the expected sojourn time, which we refer to as 
$S_{e^*,n}$, is then
\begin{align*}
S_{e^*,p} & = \frac{\lambda 2\sqrt{T} K_1(2\sqrt{T}) + 1 - \lambda}
{(1-\lambda)(1-\lambda (1 - 2T K_2(2\sqrt{T})))}.     
\end{align*}
We again here have the relation
\begin{align*}
\lambda S^*_{e,p} & = S^*_{e,n} - 1,
\end{align*}
showing that preemption is always helpful in this setting.  

For the Weibull model, 
\begin{align*}
S_{w^*,p} & = \frac{1 - \lambda + 3\lambda \sqrt{\frac{T}{2\pi}} G_{0,3}^{3,0}\left(\frac{T}{2}~|~-\frac{1}{2},0,\frac{1}{2} \right )}
{(1-\lambda)  \left ( 1- \lambda \left ( 1 - \sqrt{\frac{T^3}{2\pi}} G_{0,3}^{3,0}\left(\frac{T}{2}~|~-\frac{3}{2},0,\frac{1}{2} \right ) \right ) \right )}.
\end{align*}

\section{One-Bit Advice with Multiple Queues}
\label{sec:multiple}

In this section, we consider one-bit threshold schemes for setting
with multiple queues.  In particular, we consider the ``power of $d$
choices'' (also known as the ``balanced allocations'') setting, where
we consider the number of queues growing to infinity, and each arrival
chooses the best queue from a small constant-sized subset of randomly
selected queues \cite{DBLP:journals/siamcomp/AzarBKU99,DBLP:journals/tpds/Mitzenmacher01,vvedenskaya1996queueing}.  An advantage here of queues based on one bit of
advice is their state can easily represented, allowing the type
of mean-field analysis that is typical for such systems.
We note that analysis of the power of $d$ choices with for example
exact job sizes using queueing schemes such as Shortest Remaining Processing
Time remains an intriguing open question (see e.g. \cite{mitzenmacher2019supermarket} for more discussion),
although simply using FIFO queues and choosing the least loaded queue from
a constant number of choices has been analyzed \cite{DBLP:journals/pomacs/HellemansH18}.  

As our purpose here is primarily to demonstrate how schemes utilizing
one bit can be analyzed in this framework, we choose a relatively
simple example, based on the anecdotal 80-20 rule, that 20\% of the
jobs cause 80\% of the work.  We assume that there are two types of
jobs: long jobs have exponentially distributed service times with mean
$\mu_1$, and short jobs have exponentially distributed service times
with mean $\mu_2 < \mu_1$.  Long jobs arrive with rate $\lambda_1 n$
and short jobs arrive with rate $\lambda_2 n$, where $n$ is the number
of queues in the system.  While this model is general, it can encompass
the 80-20 rule, where long jobs are less frequent but require much more work;
for example, if $\lambda_1 \mu_1 = 4 \lambda_2 \mu_2$, then long jobs 
are relatively rare but contribute 80\% of the work.  

In the prediction setting, we assume long jobs are misclassified as
short jobs independently with probability $q_1$, and short jobs are
misclassified as long jobs independently with probability $q_2$.  (We
may view the case without predictions, where the one bit of advice is
accurate, as corresponding to $q_1 = q_2 =0$, with the resulting
equations.)  A more useful interpretation for our analysis is that a
job that is classified as a long job, or labeled long, is actually long with probability
$p_L = \lambda_1 (1-q_1)/ (\lambda_1 (1-q_1) + \lambda_2 q_2)$, and
similarly a job that is classified as a short job, or labeled short, is actually short
with probability $p_S = \lambda_2 (1-q_2)/ (\lambda_2 (1-q_2)
+ \lambda_1 q_1)$.  Similarly, the arrival rate for jobs that labeled
long is $\lambda_L = \lambda_1 (1-q_1) + \lambda_2
q_2$ and the arrival rate for jobs that are labeled short is
$\lambda_S = \lambda_2 (1-q_2) + \lambda_1 q_1$.

For serving jobs, we give labeled short jobs priority, and serve them in
FIFO fashion; similarly, labeled longs jobs are served using FIFO.
We suggest a simple, convenient method for choosing queues,
although many variations are possible and can be studied similarly.
We choose the ``best'' of $d$ queues chosen independently and
uniformly at random for a constant $d$, where we determine the best as
follows.  First, an empty queue has highest priority; an empty
queue will always be selected if it is one of the $d$ chosen.
Otherwise, we ignore the label and the time already spent being served
of the job being served.  Jobs that are predicted to be short shall
choose the queue with the fewest queued labeled short jobs, breaking
ties in favor of the queue with the fewest labeled long jobs (and
then randomly if two queues match), and similarly jobs that are
predicted to be long shall choose the queue with the smallest number
of queued labeled long jobs, breaking ties in favor of the queue with
fewer short jobs (and then randomly if two queues match).  Again, one
could imagine more complex policies based on minimizing the expected
time until service; such policies can be studied using the same
framework.

We derive equations describing the system state in the mean
field limit, where the number of queues goes to infinity.
(This approach can be formalized using the theory of density
dependent jump Markov chains, following the work of Kurtz;
see \cite{EK,Kurtz,Wormald} for examples.)
The state of a single queue can be represented by a triple $(s,\ell,c)$, where
$s$ is the number of jobs that are labeled short, $\ell$ is
the number of jobs that are labeled long, and $c$ is 1 if the
current running job is long and 2 if it is short.  The state $(0,0,0)$
is used for an empty queue;  this is the only state where $c \neq 1, 2$.
Let $x_{(s,\ell,c)}(t)$ represent the fraction of queues
in state $(s,\ell,c)$ at time $t$; we drop the $t$ where the meaning
is clear.  We use ${\hat{x}}{(s,\ell,c)}$ to refer to the equilibrium values
for these quantities.  

Note that our setting allows a relatively simple analysis by having service times be
exponentially distributed and predictions depend only on the type of
the job, instead of its running time.  Because of this, to keep the
state of a queue it suffices to keep the type of the running job, as
this gives the distribution of the remaining time it is in service.
This approach can be extended to more general service times and
predictions; see
e.g. \cite{aghajani2015mean,DBLP:journals/pomacs/HellemansH18}, for
example, for the appropriate framework.  At a high level, for such
generalizations, the state of the queue must track how long the
current running job has been in the system; the distributions function
for remaining time in service, which is derived by taking the proper
weighted average over types, then determines whether the service will
complete over the next time interval $dt$.

To write the equations describing the limiting behavior of these
systems, we use some additional notation.  Let $z_{(2,s,\ell)}$ be the
fraction of queues with lower priority over   a
queue with $s$ queued labeled short jobs and $\ell$ queue labeled
large jobs when a job labeled short arrives (in terms of being chosen by
our algorithm), and similarly 
define $z_{(1,s,\ell)}$ for when a job labeled long arrives.
Here again we drop the implicit dependence on $t$.
These $z$ values can be readily computed by dynamic programming
or even brute force given the $x_{(s,\ell,c)}$.
For
$c = 1,2$ and $c' = 1,2$, let
\begin{align*}
w_{(c',s,\ell,c)} & =  \left ( \left(z_{(c',s,\ell)} + x_{(s,\ell,1)} + x_{(s,\ell,2)} \right )^d - \left(z_{(c',s,\ell)}\right)^d \right ) \frac{x_{(s,\ell,c)}}{x_{(s,\ell,1)} + x_{(s,\ell,2)}}.
\end{align*}
Then $w_{(c',s,\ell,c)}$ gives the probability that an incoming job
labeled $c'$ chooses a queue in
state $(s,\ell,c)$.  
For the empty queue, we have the special case
\begin{align*}
w_{(c',0,0,0)} & = 1 - \left(1-x_{(0,0,0)}\right )^d,
\end{align*}
and is it convenient to let $w_{(c',s,\ell,c)} = 0$ if $s <0 $ or $\ell < 0$.

The limiting mean field equations when $s >0$ are then
\begin{align*}
\frac{dx_{(s,\ell,1)}}{dt} & = \lambda_S w_{(2,s-1,\ell,1)}
+ \lambda_L w_{(1,s,\ell-1,1)} + \mu_1 x_{(s+1,\ell,1)}(1-p_S) + \mu_2 x_{(s+1,\ell,2)}(1-p_S)  \\
& - \left (\mu_1 x_{(s,\ell,1)} + \lambda_S w_{(2,s,\ell,1)}
+ \lambda_L w_{(1,s,\ell,1)} \right),
\end{align*}
and 
\begin{align*}
\frac{dx_{(s,\ell,2)}}{dt} & = \lambda_S w_{(2,s-1,\ell,2)}
+ \lambda_L w_{(1,s,\ell-1,2)} + \mu_1 x_{(s+1,\ell,1)}p_S + \mu_2 x_{(s+1,\ell,2)}p_S  \\
& - \left (\mu_2 x_{(s,\ell,2)} + \lambda_S w_{(2,s,\ell,2)}
+ \lambda_L w_{(1,s,\ell,2)}
\right).
\end{align*}

The cases where $s = 0$ and $\ell > 0$ are given by
\begin{align*}
\frac{dx_{(0,\ell,1)}}{dt} & = \lambda_L w_{(1,0,\ell-1,1)} + \mu_1 x_{(1,\ell,1)}(1-p_S) + \mu_2 x_{(1,\ell,2)}(1-p_S) + \mu_1 x_{(0,\ell+1,1)}p_L
+ \mu_2 x_{(0,\ell+1,2)}p_L
  \\
& - \left (\mu_1 x_{(0,\ell,1)} + + \lambda_S w_{(2,0,\ell,1)}
+ \lambda_L w_{(1,0,\ell,1)} \right),
\end{align*}
and 
\begin{align*}
\frac{dx_{(0,\ell,2)}}{dt} & = \lambda_L w_{(1,0,l-1,2} + \mu_1 x_{(1,\ell,1)}p_S + \mu_2 x_{(1,\ell,2)}p_S + \mu_1 x_{(0,\ell+1,1)}(1-p_L)
+ \mu_2 x_{(0,\ell+1,2)}(1-p_L) \\
& - \left (\mu_2 x_{(0,\ell,2)} + 
+ \lambda_S w_{(2,0,\ell,2)}
+ \lambda_L w_{(1,0,\ell,2)}
\right),
\end{align*}

And finally, for queues without any waiting jobs, we have
\begin{align*}
\frac{dx_{(0,0,1)}}{dt} & = \lambda_L p_L w_{(1,0,0,0)} +
\lambda_S (1-p_S) w_{(1,0,0,0)} \\
& + \mu_1 x_{(0,1,1)} p_L 
+ \mu_2 x_{(0,1,2)} p_L 
+ \mu_1 x_{(1,0,1)} (1-p_S)
+ \mu_2 x_{(1,0,2)} (1-p_S) \\
& - \left (\mu_1 x_{0,0,1}
+ \lambda_S w_{(2,0,0,1)}
+ \lambda_L w_{(1,0,0,1)}\right );
\end{align*}

\begin{align*}
\frac{dx_{(0,0,2)}}{dt} & = \lambda_S p_S w_{(2,0,0,0)} +
\lambda_L (1-p_L) w_{(2,0,0,0)} \\
& + \mu_1 x_{(0,1,1)} (1-p_L)
+ \mu_2 x_{(0,1,2)} (1-p_L)
+ \mu_1 x_{(1,0,1)} p_S
+ \mu_2 x_{(1,0,2)} p_S \\
& - \left (\mu_1 x_{0,0,2} 
+ \lambda_S w_{(2,0,0,2)}
+ \lambda_L w_{(1,0,0,2)} \right);
\end{align*}

\begin{align*}
\frac{dx_{(0,0,0)}}{dt} & = \mu_1 x_{(0,0,1)} + \mu_2 x_{(0,0,2)} - 
\left(\lambda_S w_{(2,0,0,0)}
+ \lambda_L w_{(1,0,0,0)}\right ).
\end{align*}

In section~\ref{sec:results} below, we compare the results from calculating the differential equations results numerically with simulations.

\section{Simulations and Numerical Results}
\label{sec:results}

\subsection{Single Queues}
\label{subsec:results1}

In the simulations for single queues, each data point is obtained by
simulating initially empty queues over 1000000 units of time, and
taking the average response time for all jobs that terminate after
time 100000.  We then take the average of over 100 simulations.
Waiting for the first 10\% allows the system to approach the
stationary distribution, and we run for sufficient time that recording
only completed jobs has small influence.

Before presenting results, we emphasize that we have checked the
experimental results for single queues against the equations we have derived in
sections~\ref{sec:thresh} and~\ref{sec:predict}.  They match very closely; in general terms,
nearly all the averaged simulation results presented are within 1\% of
the values derived from the equations.  (Individual simulation runs
can vary more significantly; the maximum and minimum times over our trials
vary by 5-10\% for exponentially distributed service times, and by 10-20\% for
Weibull service times.)  As such, we do not present further results
comparing equations to simulations here.\footnote{While arguably we could have
simply trusted the equations, we find verifying results via simulation
worthwhile.}  

Our first simulations are for exponentially distributed service times.
While we have done more simulations at various
arrival rates, we present results for $\lambda = 0.8, 0.9,$ and $0.95$.
This focuses on the more interesting case of reasonably high arrival
rates, while keeping the values within a reasonable range for presentation.
As a baseline, when the arrival rate is $\lambda$, the expected time a
job spends in such a system in equilibrium with a FIFO queue is
$1/(1-\lambda)$.  

\begin{figure}
    \centering
    \begin{minipage}{0.45\textwidth}
        \centering
        \includegraphics[width=0.9\textwidth]{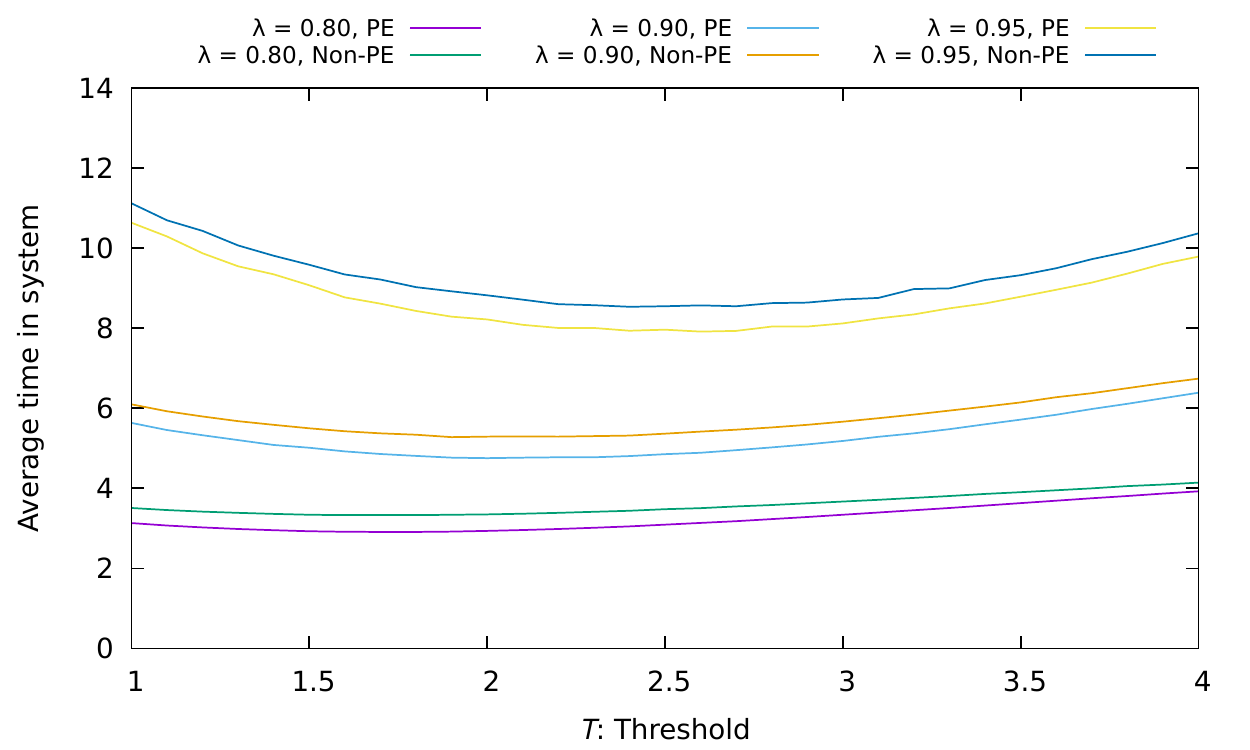}     
        \caption{Performance under threshold schemes with exact information, exponentially distributed service times.}
    \label{fig:exact.pdf}
    \end{minipage}\hfill
    \begin{minipage}{0.45\textwidth}
        \centering
        \includegraphics[width=0.9\textwidth]{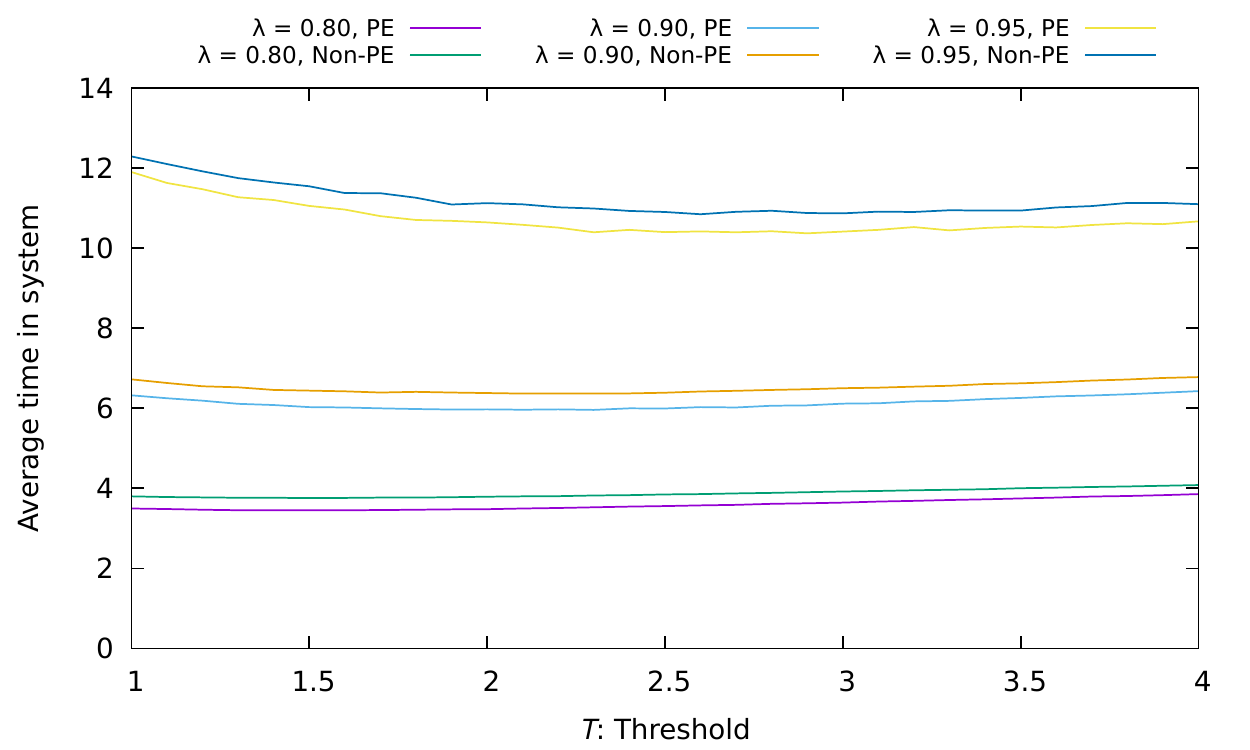} 
        \caption{Performance under threshold schemes with predicted information, 
exponentially distributed service times, exponential predictions.}
    \label{fig:predicted.pdf}
    \end{minipage}\hfill
\end{figure}

We first show the results of the experiments, comparing the results
with and without preemption, and with and without prediction.
Figure~\ref{fig:exact.pdf} shows the results as the threshold varies given correct
one-bit advice, and Figure~\ref{fig:predicted.pdf} shows the results under the
exponential prediction model. The figures are at the same scale so
results can be compared.  We see here that preemption, as expected,
provides some gains, and the cost for using prediction is not too
large.  In particular, one bit of even sometimes incorrect advice
substantially reduces the average time in system over simple FIFO
queueing.  The results show that in this setting choosing a threshold
near the optimal rather than the optimal does not substantially affect
the results.  

We also do experiments for the Weibull distribution with cumulative
distribution $1-e^{-\sqrt{2x}}$.  As a baseline, when the arrival
rate is $\lambda$, the expected time a job spends in such a system in
equilibrium with a FIFO queue under this Weibull distribution is
$(1+\lambda)/(1-\lambda)$.

\begin{figure}
    \centering
    \begin{minipage}{0.45\textwidth}
        \centering
        \includegraphics[width=0.9\textwidth]{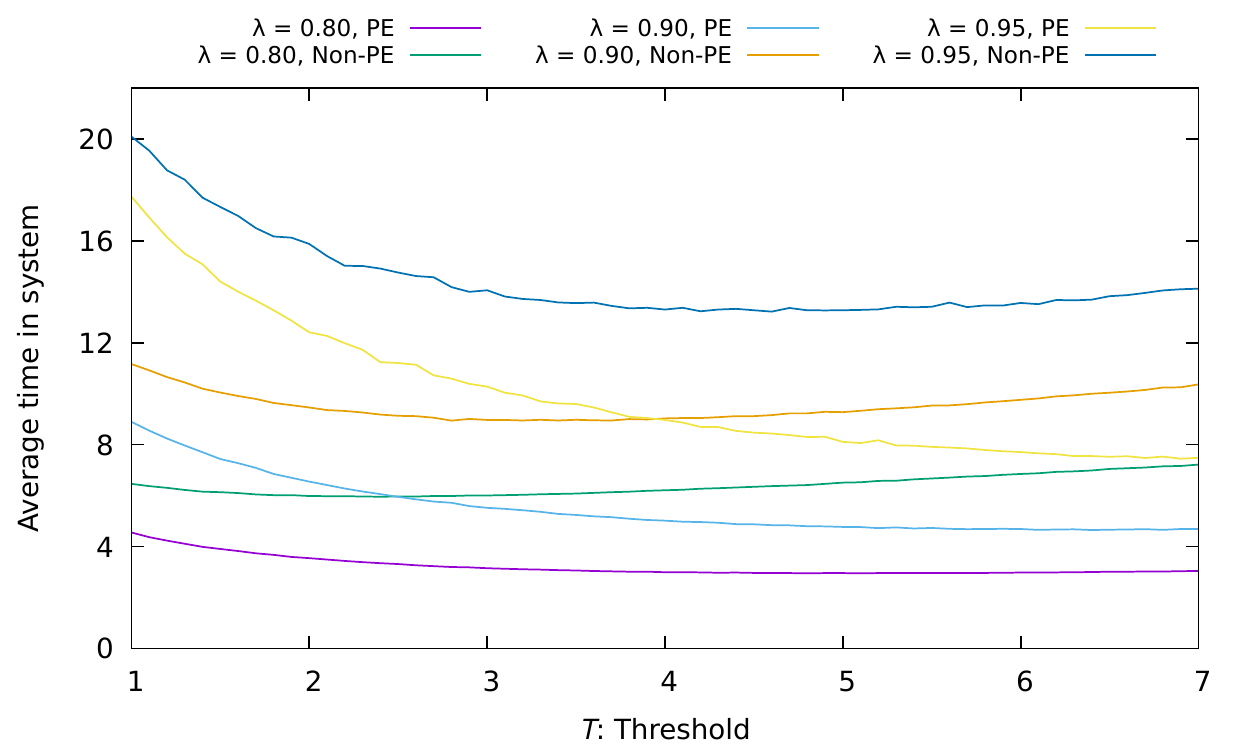}     
        \caption{Performance under threshold schemes with exact information, Weibull distributed service times.}
    \label{fig:exact.ht.pdf}
    \end{minipage}\hfill
    \begin{minipage}{0.45\textwidth}
        \centering
        \includegraphics[width=0.9\textwidth]{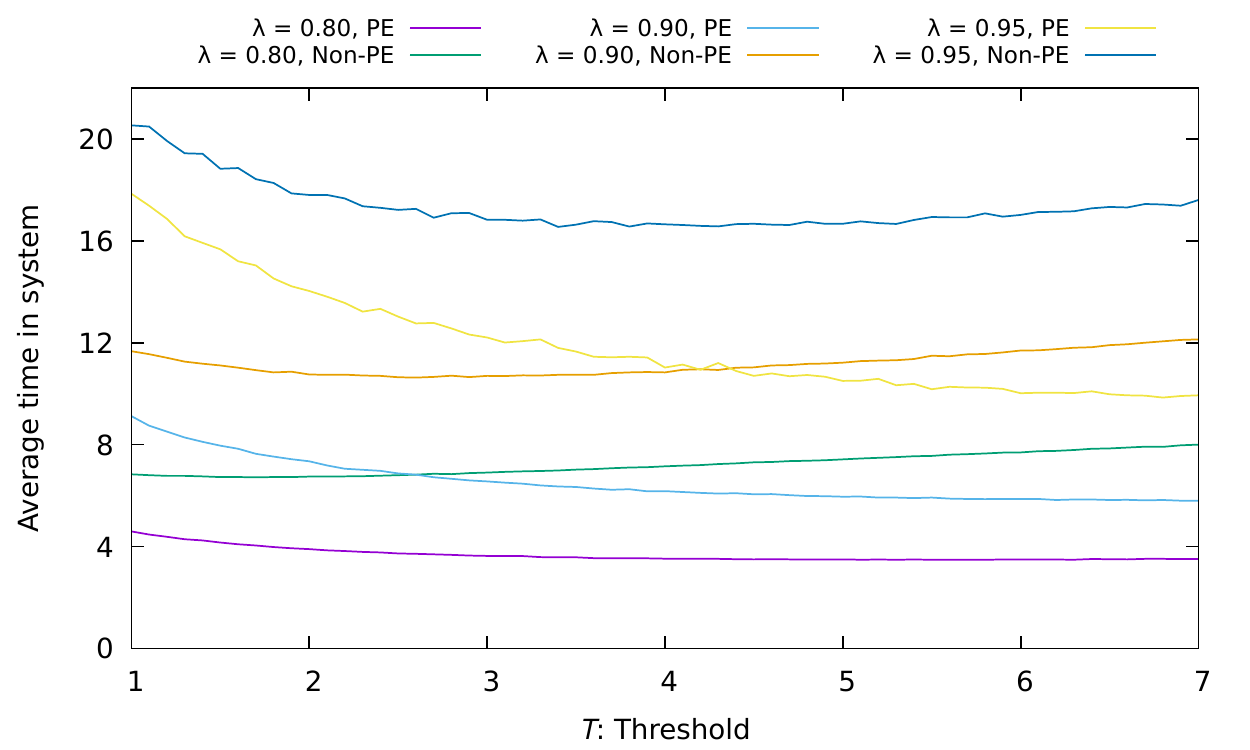} 
        \caption{Performance under threshold schemes with predicted information, Weibull distributed service times, exponential predictions.}
    \label{fig:predicted.ht.pdf}
    \end{minipage}\hfill
\end{figure}

Similar to before, Figure~\ref{fig:exact.ht.pdf} shows the results as the threshold
varies given correct one-bit advice, and Figure~\ref{fig:predicted.ht.pdf} shows the
results under the exponential prediction model.  Note that the range
of thresholds is much larger.  One would expect larger thresholds
would be optimal for a heavy-tailed distribution, as the downside of
having a very large job at the front of the queue is more substantial.
Further, in this setting, preemption offers more substantial gains,
and clearly pushes the optimal threshold to larger values, as
preemption significantly reduces the impact of a long job holding the
queue.  While the cost of using predicted advice over optimal advice
is larger, the potential (and actual) gains from using predicted
advice are even more substantial. 

To shed additional light, we compare one-bit advice to no advice, in
which case the queue uses FIFO, and full knowledge of the processing
times, in which case the queue uses Shortest Remaining Processing Time
(SRPT).  In the case of predictions, we consider the exponential
prediction model, comparing our schemes against FIFO and Shortest
Predicted Remaining Processing Time \cite{mitz2019scheduling}, which
uses SRPT scheduling on the predicted times. For FIFO results we
simply use the standard formula for expected time in the system; we
could similarly use formulae for the other results, but present
results from simulations.  In particular, for our one-bit schemes, we
choose the best threshold from simulation result, and allow
preemptions.  As we can see in Tables~\ref{tab:table-exp} and \ref{tab:table-ht}, one-bit schemes greatly improve on FIFO
across the board, with a greater improvement for Weibull
distributions, as one would expect.  Indeed, one-bit threshold schemes
achieve a large fraction of the benefit that would arise from full
knowledge of processing times, and one-bit predictions achieve a large
fraction of the benefit that would arise from more detailed
predictions.  We also find that preemption is helpful, and moreso for
the heavier tailed distributions, as one would expect.  Perhaps most
important, using predictions provides large gains, nearly as good as
with exact information, showing the large potential for even simple
predictions to provide large value in scheduling.  

{\footnotesize
\begin{table}
\begin{center}
\begin{tabular}{|c||c||c|c|c||c|c|c|}
\hline
           & FIFO & THRESHOLD  & THRESHOLD   & SRPT & PREDICTION & PREDICTION & SPRPT \\
$\lambda$  &      & NO PREEMPT & PREEMPT &      & NO PREEMPT  & PREEMPT &  \\ [0.5ex]
 \hline\hline
0.50	 & 	2.000	 & 	1.783	 & 	1.564	 & 	1.425	 & 	1.850	 & 	1.698	 & 	1.659 \\ \hline
0.60	 & 	2.500	 & 	2.089	 & 	1.814	 & 	1.604	 & 	2.209	 & 	2.013	 & 	1.940 \\ \hline
0.70	 & 	3.333	 & 	2.542	 & 	2.203	 & 	1.875	 & 	2.761	 & 	2.517	 & 	2.369 \\ \hline
0.80	 & 	5.000	 & 	3.329	 & 	2.910	 & 	2.355    & 	3.757	 & 	3.451	 & 	3.143 \\ \hline  
0.90	 & 	10.00	 & 	5.278	 & 	4.755	 & 	3.552	 & 	6.366	 & 	5.960	 & 	5.097 \\ \hline
0.95	 & 	20.00	 & 	8.535	 & 	7.914	 & 	5.532	 & 	10.848	 & 	10.372	 & 	8.424 \\ \hline
0.98	 & 	50.00	 & 	16.495	 & 	15.735	 & 	10.436	 & 	22.418	 & 	21.909	 & 	16.696 \\ \hline
\end{tabular}
\caption{Results for exponentially distributed service times.  Prediction
results are using exponential predictions.} 
\label{tab:table-exp}
\end{center}
\end{table}
}

{\footnotesize
\begin{table}
\begin{center}
\begin{tabular}{|c||c||c|c|c||c|c|c|}
\hline
 & FIFO & THRESHOLD      & THRESHOLD   & SRPT & PREDICTION & PREDICTION & SPRPT \\
 &      & NO PREEMPT & PREEMPT &      & NO PREEMPT  & PREEMPT &   \\ [0.5ex]
 \hline\hline
0.50     &      4.000    &      3.012   &      1.608  &      1.411  &      3.155  &      1.736  &      1.940 \\ \hline
0.60     &      5.500    &      3.676  &      1.867  &      1.574  &      3.918  &      2.062  &      2.280 \\ \hline
0.70     &      8.000    &      4.565  &      2.258  &      1.813  &      4.983  &      2.568  &      2.750 \\ \hline
0.80     &      13.00    &      5.955  &      2.951  &      2.217    &      6.721  &      3.481  &      3.519 \\ \hline
0.90     &      29.00    &      8.940  &      4.649  &      3.154  &      10.630  &      5.790  &      5.224 \\ \hline
0.95     &      58.00    &      13.223  &      7.448  &      4.517  &      16.546  &      9.846   &      7.788 \\ \hline
0.98     &      148.0    &      22.451  &      15.194   &      7.666  &      29.346  &      20.918  &      13.404 \\ \hline
\end{tabular}
\caption{Results for Weibull distributed service times.  Prediction
results are using exponential predictions.} 
\label{tab:table-ht}
\end{center}
\end{table}
}

\subsection{Multiple Queues}
\label{subsec:results2}

We present results here for an example using the power of two choices,
to demonstrate the results from differential equations match
simulations, and to show the effectiveness of working with
predictions.  In our example we follow the 80-20 rule; we choose
parameters $\lambda_1 = 0.225$, $\mu_1 = 3.2$, $\lambda_2 = 0.90$, and
$\mu_2 = 0.20$.  The overall load on the system is therefore $0.9$.
For simulations, each data point is obtained by simulating systems of
1000 (initially empty) queues over 100000 units of time, and taking
the average response time for all jobs that terminate after time
10000.  We then take the average of over 100 simulations.  For the
differential equations, we simply used Euler's method over times steps
of $10^{-5}$ over time $10^4$.  (This provides an accurate calculation
for the ``fixed point'', or stationary distribution corresponding to
the solution of these equations.)  All experiments use two choices.  We provide simulation results for
randomly choosing a single queue and using FIFO, choosing a queue based
on the least loaded and using SRPT within the queue, and choosing the shorter
of two queues and FIFO processing for comparison.  We provide simulation
results for various predictions, where the two values after ``Pred'' in the table are $q_1$ (misprediction for long jobs) and $q_2$ (short jobs), respectively.  

{\footnotesize
\begin{table}
\begin{center}
\begin{tabular}{|c||c|c|}
\hline
 &  Simulations & Diff. Eqns. \\ [0.5ex]
 \hline\hline
1 Choice    & 24.208  & $---$ \\ \hline
SRPT        &  2.366    & $---$ \\ \hline
Shorter Queue, FIFO  & 4.967  & $---$ \\ \hline  
Pred 0.0, 0.0  & 3.394  & 3.392\\ \hline  
Pred 0.1, 0.1  & 3.690  & 3.688 \\ \hline
Pred 0.2, 0.2 & 4.010  & 4.007  \\ \hline
Pred 0.3, 0.3   & 4.353  & 4.347 \\ \hline   
Pred 0.4, 0.4   & 4.717  & 4.711 \\ \hline   
Pred 0.5, 0.5   & 5.105  & 5.098 \\ \hline   
Pred 0.2, 0.4   & 4.280  & 4.276 \\ \hline   
Pred 0.4, 0.2   & 4.402  & 4.395 \\ \hline   
Pred 0.11, 0.61   &  4.617 & 4.611 \\ \hline   
\end{tabular}
\caption{Results for queueing systems with 1000 queues, 2 choices and baseline comparisons.} 
\label{tab:table-qs}
\end{center}
\end{table}
}

The main takeaways from Table~\ref{tab:table-qs}, beyond the 
fact that the differential equations are quite accurate (less than 0.2\% difference in these examples), are that one-bit predictions can provide benefits over the already excellent performance of choosing the shorter queue; even when all predictions are only 60\% accurate, we seem some gain in performance.  Considering $q_1 = 0.2$ and $q_2 = 0.4$ along with $q_1 =0.4$ and $q_2 = 0.2$ shows that predictions for long jobs are more important, even though there are much fewer long jobs.  The effect of giving a long job higher priority, where it can block short jobs, has a more prominent effect than misclassifying a short job.  This demonstrates that the goal of a machine learning algorithm in this setting should not be simply to maximize the number of correct predictions;  a machine learning algorithm can do better by predicting the long jobs well. (See \cite{mitzenmacher2019supermarket} for a similar discussion.)  

As an extreme example of this, choosing $q_1 = 0.11$ and $q_2 = 0.61$ leads to a total error rate of 51\% over all jobs, as short jobs have a much higher arrival rate than long jobs.  But even though more than half the jobs types are predicted incorrectly, because long jobs are predicted correctly most of the time, such predictions still perform notably better than not using predictions and just choosing the shorter queue.  

\section{Conclusions}

We have looked at the setting of queueing systems with one bit of advice, where a primary motivation is the potential for machine learning algorithms to provide simple but useful predictions to improve scheduling.  In the case of single queues, we see that a natural probabilistic model for predictions leads to relatively straightforward equations that can be used to determine where one would ideally choose a threshold to separate long and short jobs.  For large-scale queueing systems, where the power of two choices can be used, we have shown that one-bit prediction can allow for fluid limit analysis.  We view this as a potential step forward for the interesting open problem of determining the behavior of systems using the power of two choices with scheduling via shortest remaining processing time or other scheduling schemes based on the service time.   

We believe there remain several interesting directions to explore in this space.  The use of predictions in more complex settings, such as call centers, may provide significant value.  A challenging underlying questions, when ``jobs'' may correspond to people, is how to define appropriate notions of fairness, so tht jobs that are mispredicted by a machine learning algorithm do not suffer overly from the algorithm's behavior.

\end{document}